\documentclass{webofc}

\usepackage[varg]{txfonts}   
\usepackage{hyperref}
\usepackage{url}
\hypersetup{colorlinks=true,citecolor=blue,urlcolor=blue,linkcolor=blue}
\begin{document}
\title{\mbox {Recent results on heavy flavours and quarkonia from ALICE}}
%
%

\author{\firstname{Fiorella Fionda}\inst{1}\fnsep\thanks{\email{fiorella.fionda@cern.ch}}\lastname{on behalf of the ALICE Collaboration} 
}

\institute{University and INFN, Cagliari (Italy)
          }

\abstract{Heavy-flavour hadrons, containing at least one charm or beauty quark, are excellent probes of the deconfined medium created in ultra-relativistic heavy-ion collisions, known as quark--gluon plasma. Results in smaller collision systems, such as proton--proton and p--Pb collisions, besides representing an important baseline for interpreting heavy-ion measurements, are crucial to test perturbative QCD calculations and hadronisation mechanisms in the absence of hot medium effects, as well as to search for commonalities with heavy-ion systems. Recently, measurements in proton--proton and p--Pb collisions have revealed unforeseen features with respect to the expectations based on previous results from ${\rm e}^{+}{\rm e}^{-}$ and ep collisions, showing that fragmentation fractions of heavy quarks are not universal. In this contribution, an overview of the most recent ALICE heavy-flavour measurements, along with the comparison to available calculations, will be discussed. 
}
\maketitle
\section{Introduction}
\label{intro}
Quarkonia, bound states of a $\rm c \overline{\rm c}$ or a $\rm b \overline{\rm b}$ pair, and open heavy-flavour (HF) hadrons, 
which have non-zero charm or beauty quantum numbers, are effective probes of several aspects of quantum chromodynamics (QCD). 
Comprehensive measurements of quarkonia and open HF hadrons in Pb--Pb collisions allow for testing the mechanisms of heavy-quark transport, energy loss, and coalescence effects during hadronisation in the presence of the quark--gluon plasma (QGP). In order to correctly interpret results in Pb--Pb collisions, measurements in proton--proton (pp) and p--Pb collisions are fundamental baselines. In particular, pp collisions provide details on the vacuum production, and studies in p--Pb collisions are useful to examine the so-called cold nuclear matter effects. 
Besides serving as reference, HF measurements in hadronic collisions provide a stringent test of QCD since the large quark masses imply high momentum transfer processes that make perturbative QCD (pQCD) calculations applicable to a large extent. According to the factorisation approach~\cite{Factorisation}, the production cross section of HF hadrons in hadronic collisions can be described as the convolution of three terms: the parton distribution functions (PDFs) of the colliding nucleons; the partonic cross-section responsible for the production of the heavy-quark pair, which can be computed using pQCD calculations; and the non-perturbative evolution into the bound states. Concerning the latter contribution, in the case of quarkonia different semi-phenomenological approaches, such as those implemented in the Color Evaporation Model (CEM) and Non-Relativistic QCD (NRQCD), are employed (see Ref.~\cite{Andronic} for a comprehensive review). For open HF hadrons, fragmentation functions, which describe the fraction of the heavy-quark momentum carried by the HF hadron, are considered. In the factorisation approach, the hadronisation process is taken to be independent of the collision system, and this assumption implies identical fragmentation fractions and fragmentation functions for ${\rm e}^{+}{\rm e}^{-}$, ep and pp collisions. However, in the last decade a large wealth of results related to HF baryon production have disproved the assumed universality of the fragmentation fractions. In particular, an enhancement was observed for the first time by the ALICE Collaboration at midrapidity in pp collisions for the prompt $\Lambda_{\rm c}^{+}$/${\rm D}^{0}$ production yield ratio compared to similar results in leptonic collisions~\cite{LambdaCEnh}, and this is even more pronounced for charm-strange baryon states production relative to ${\rm D}^{0}$ mesons~\cite{XicAndOmegaEnh}. In the beauty sector, the measurement of the $\Lambda_{\rm b}^{0}$ baryon production relative to that of beauty mesons at forward rapidity from the LHCb Collaboration~\cite{LHCb:2019fns} shows an enhancement of the corresponding fragmentation fraction in pp collisions similar to that observed for charm quarks at midrapidity. As a consequence, state-of-the-art calculations employing a model of quark fragmentation constrained by leptonic measurements, such as those considered in PYTHIA 8 with the default tune (Monash)~\cite{pythia8monash}, are challenged. Several models attempt to reproduce the observed baryon enhancement assuming either modified hadronisation in a parton-rich environment with respect to in-vacuum fragmentation via different mechanisms, such as colour reconnection beyond the leading-colour approximation (CR-BLC)~\cite{PythiaBLC} in PYTHIA 8 and coalescence~\cite{Catania,Powlang}, or enhanced baryon production originating from the decay of unobserved charm-baryon resonant states~\cite{He}. 
\newline The ALICE detector has excellent capabilities for reconstructing a large variety of HF hadrons in a wide rapidity window and down to very low transverse momentum ($p_{\rm T}$), exploiting both inclusive and exclusive decay channels. It went through a major upgrade during Long Shutdown 2 (2019-2021)~\cite{ALICEupgrade} of the LHC and restarted data taking at the beginning of LHC Run 3 in 2022. The main upgraded detector subsystems in the central barrel are the Time Projection Chamber (TPC) and the Inner Tracking System (ITS). In particular, the upgraded readout system of the TPC allows it to cope with a continuous readout mode in Run 3, while the upgraded ITS provides an approximately 3 (6) times improvement in pointing resolution in the transverse (longitudinal) direction. In addition, the Muon Forward Tracker was installed at forward rapidity, enabling the reconstruction of secondary vertices from charm- and beauty-hadron decays in the acceptance of the muon spectrometer. 

\section{Highlights from pp collisions}

\begin{figure*}
\centering
\includegraphics[width=6.4cm,clip]{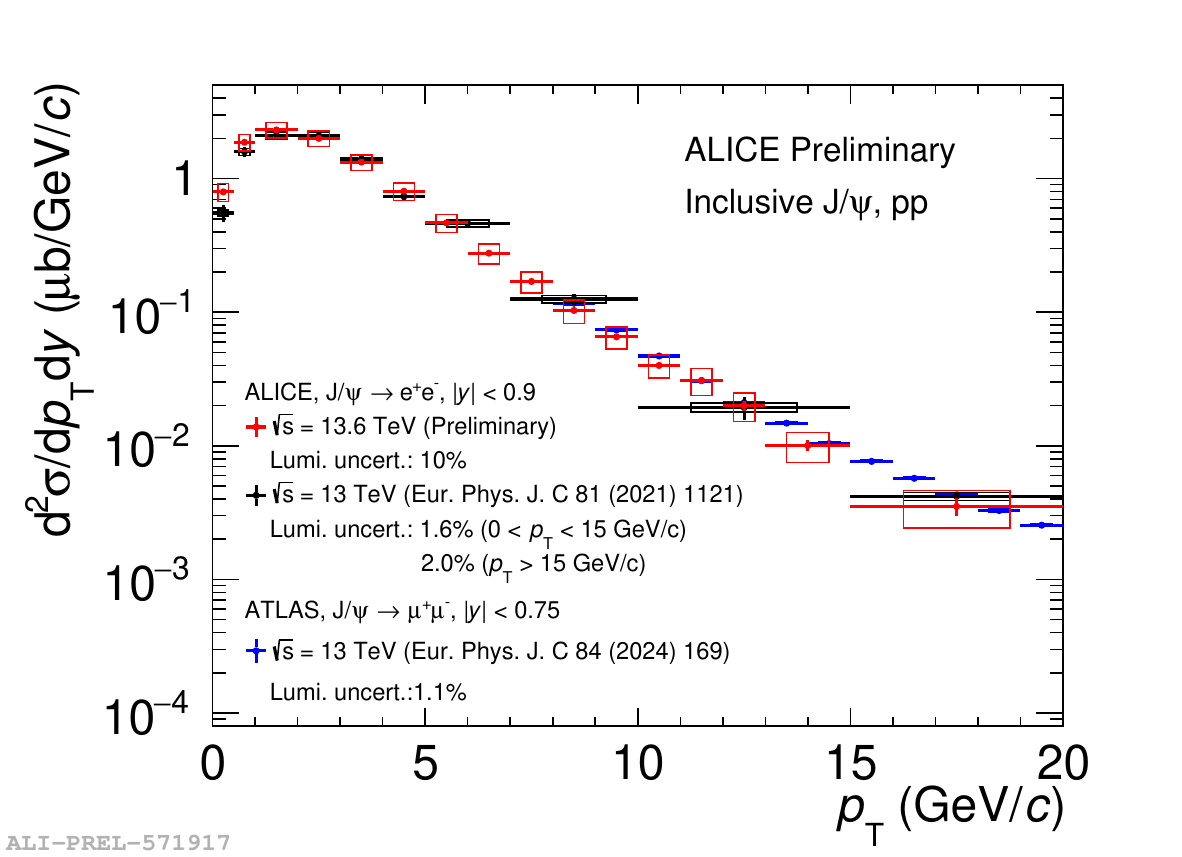}
\includegraphics[width=6.4cm,clip]{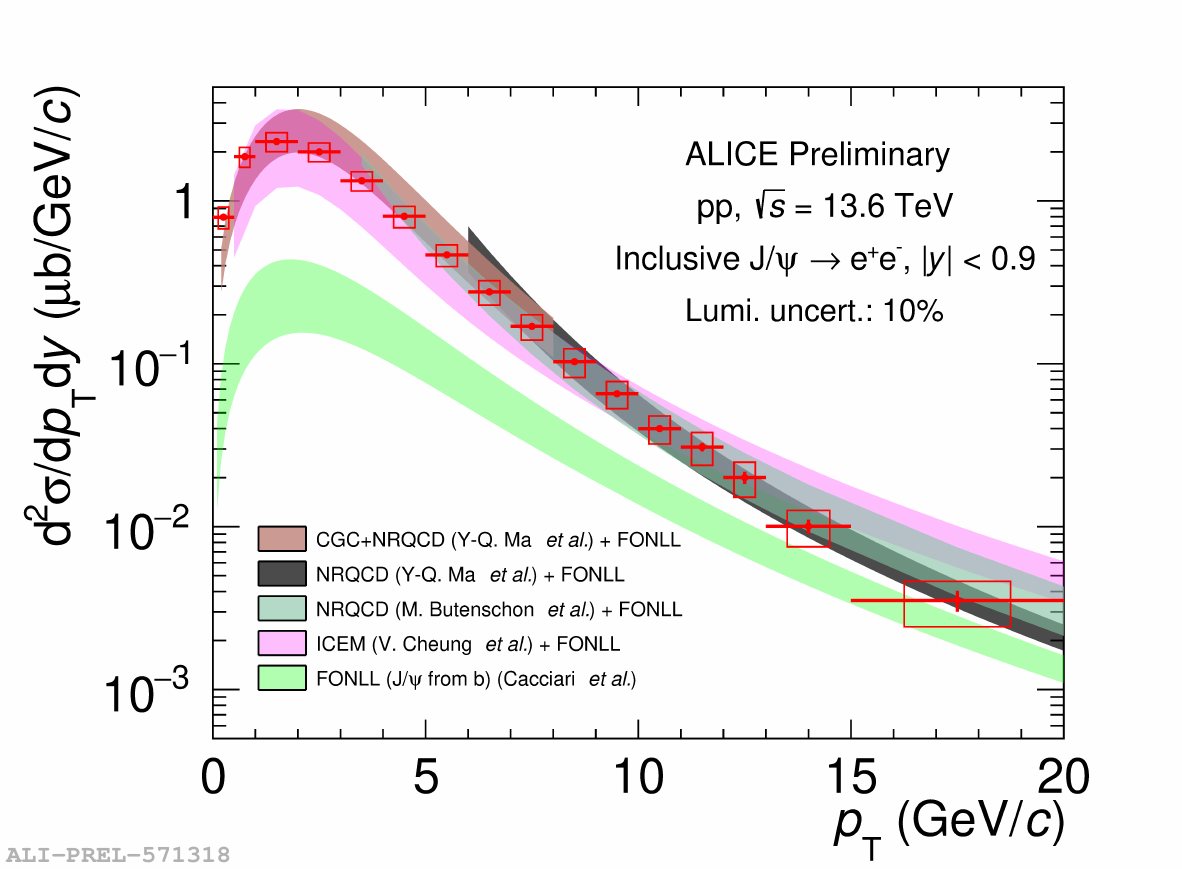}
	\caption{Inclusive J/$\psi$ cross section measured at midrapidity in pp collisions at $\sqrt{s}$ = 13.6 TeV compared to similar measurements from ALICE and ATLAS at  $\sqrt{s}$ = 13 TeV in the left panel (see references on the plot) and with several charmonium production models~\cite{Buth,Ma,Ma1,icem} coupled to FONLL~\cite{fonll} in the right panel.}
\label{JpsiCrossSec}       
\end{figure*}

\begin{figure*}
\centering
\includegraphics[width=5.2cm,clip]{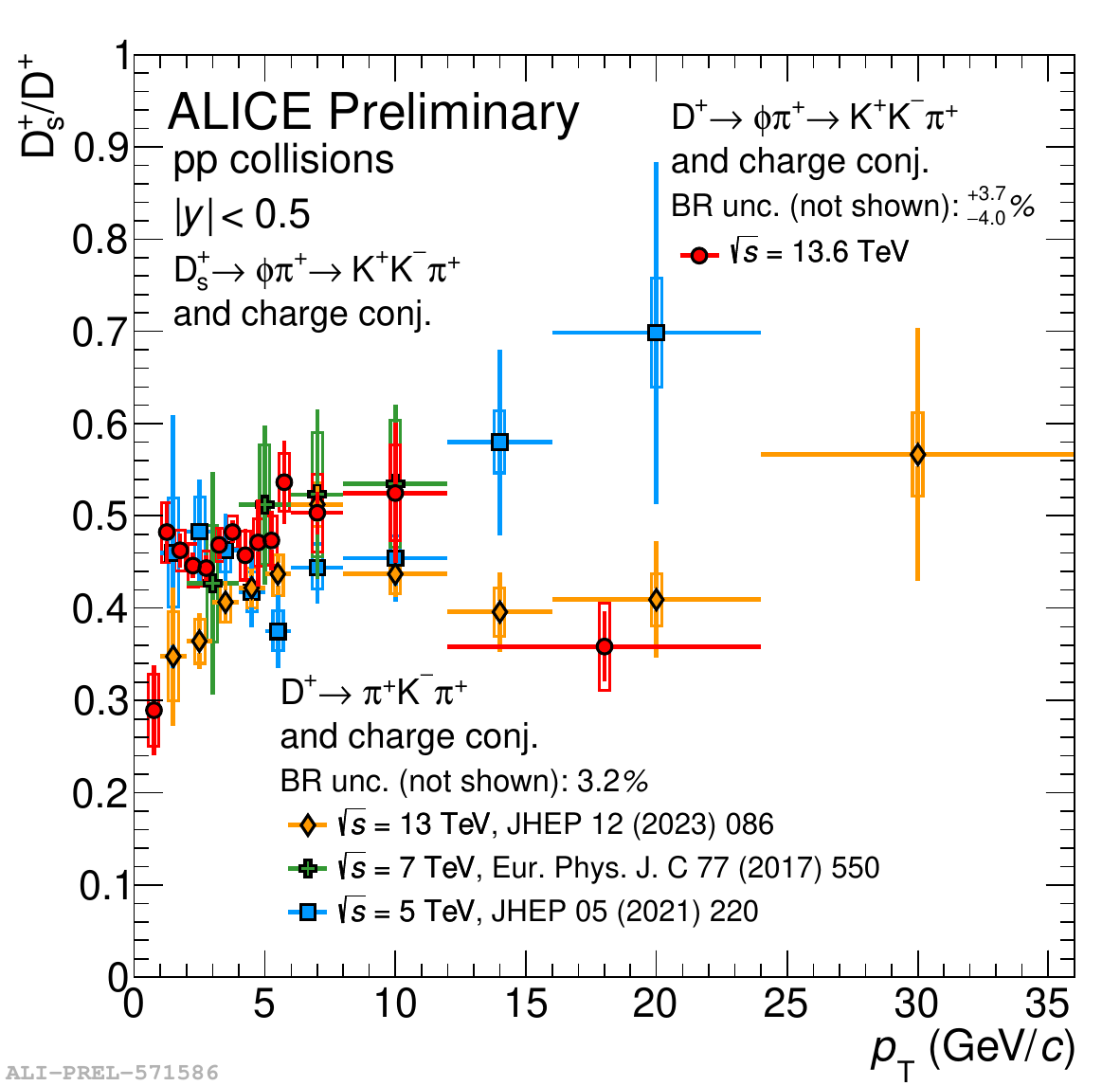}
	\hspace{0.8cm}
\includegraphics[width=5.2cm,clip]{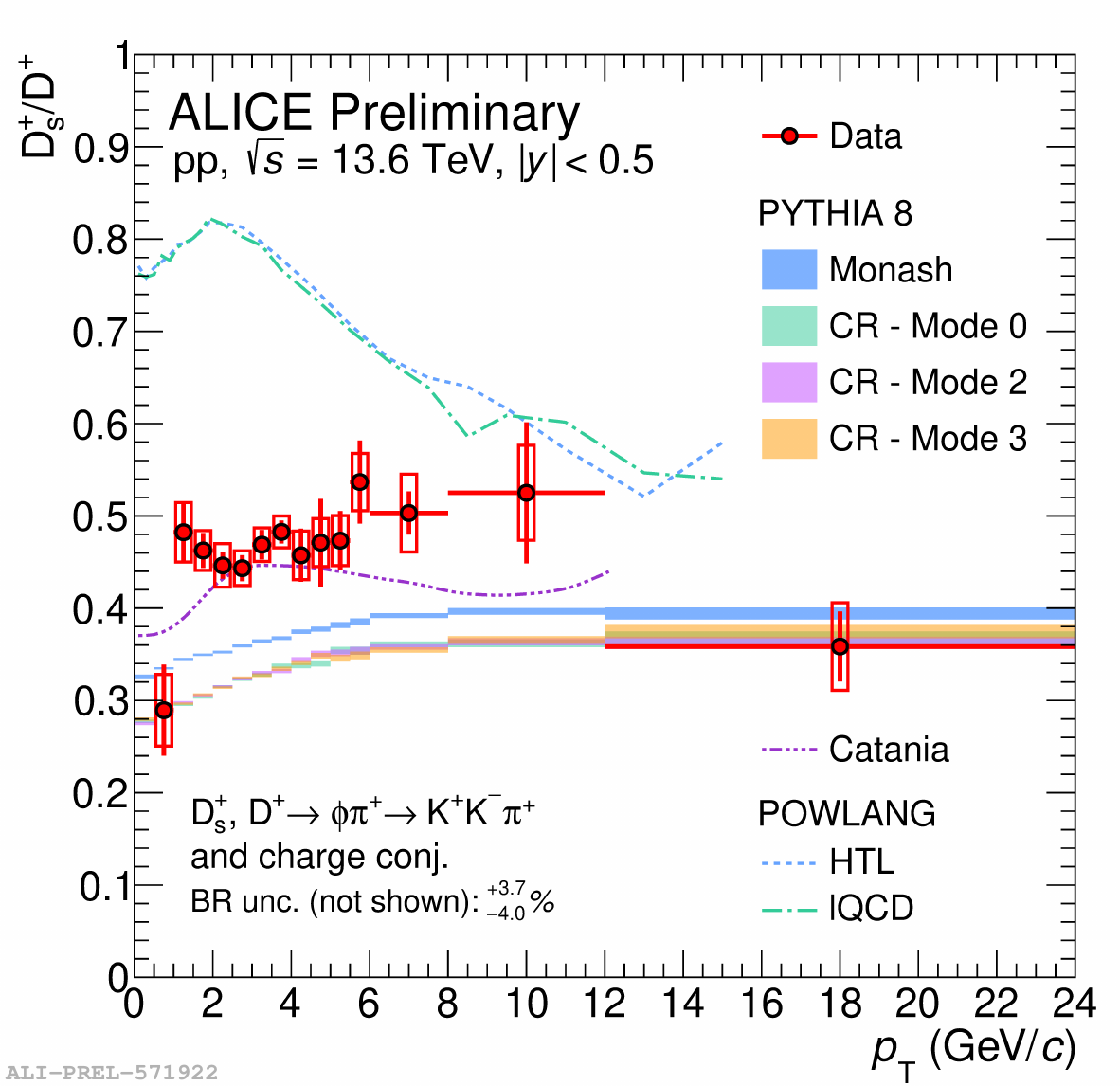}
	\caption{Prompt ${\rm D}_{\rm s}^{+}$/${\rm D}^{+}$ production yield ratio measured at midrapidity in pp collisions at $\sqrt{s}$ = 13.6 TeV compared to similar measurements at lower centre-of-mass energies from Run 2 in the left panel (see references on the plot) and with model calculations~\cite{pythia8monash,PythiaBLC,Catania,Powlang} in the right panel.   }
\label{DmesonRatios}       
\end{figure*}

Figure~\ref{JpsiCrossSec} shows the new preliminary measurement of the inclusive J/$\psi$ cross section reconstructed at midrapidity in pp collisions at $\sqrt{s}$ = 13.6 TeV. In the left panel, the results are compared with similar midrapidity Run 2 measurements from ALICE and ATLAS (see references on the left panel), the latter available for $p_{\rm T} > 7$~GeV/$c$. The comparison indicates a good agreement between Run 2 and Run 3 results, as well as a significantly improved granularity with respect to previous ALICE measurements. The cross section is described within uncertainties by NRQCD-based models~\cite{Buth,Ma,Ma1} and an improved version of the CEM~\cite{icem}, both coupled to FONLL~\cite{fonll} to account for the contribution of J/$\psi$ originating from beauty hadron decays, as shown in the right panel of Fig.~\ref{JpsiCrossSec}. The large uncertainties of the model calculations do not allow for discriminating among different models, however these measurements represent an important input for tuning such models, especially in the low-$p_{\rm T}$ region. 
\newline Preliminary measurements in the open HF sector based on Run 3 data are also carried out. The prompt ${\rm D}_{\rm s}^{+}$/$\rm D^{+}$ production yield ratio at midrapidity in pp collisions at $\sqrt{s}$ = 13.6 TeV is depicted in Fig.~\ref{DmesonRatios}. The comparison to previous Run 2 results in the left panel clearly shows that preliminary Run 3 measurements are compatible with published results at lower centre-of-mass energies.
Furthermore, they have significantly increased granularity and are extended down to a lower $p_{\rm T}$ (0.5 GeV/$c$) compared to previous measurements. In the right panel of Fig.~\ref{DmesonRatios}, the prompt ${\rm D}_{\rm s}^{+}$/$\rm D^{+}$ production yield ratio is compared to model calculations. PYTHIA 8 calculations with Monash~\cite{pythia8monash} and CR-BLC~\cite{PythiaBLC} tunes, the latter introduced to improve the description for the baryons, underpredict the data, while POWLANG~\cite{Powlang}, which assumes an expanding fireball in pp collisions and hadronisation occurring via coalescence, overshoots the measured ratio. The Catania~\cite{Catania} model, which considers both coalescence and fragmentation for the charm quark hadronisation, provides a description closer to the data. 
\begin{figure*}
\centering
\includegraphics[width=5.6cm,clip]{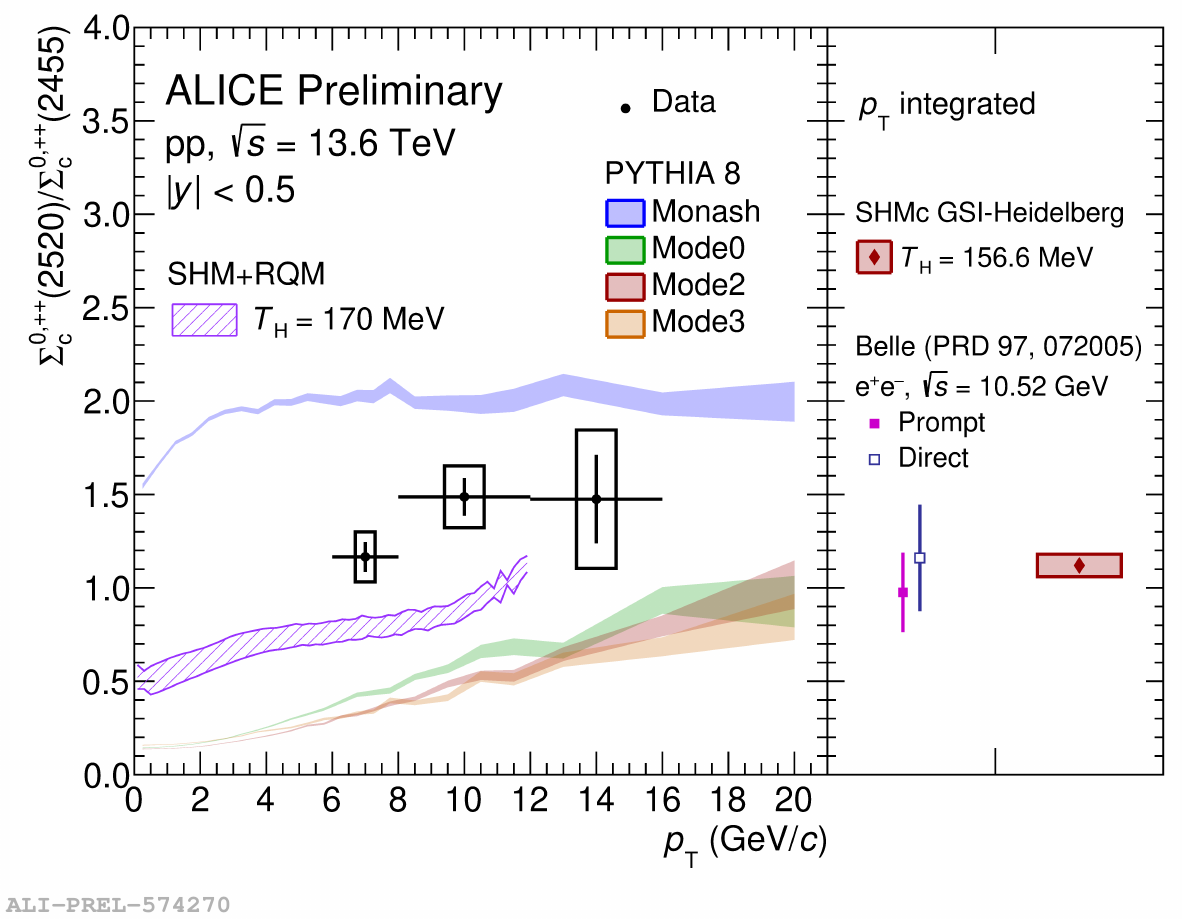}
\hspace{0.4cm}
\includegraphics[width=5.6cm,clip]{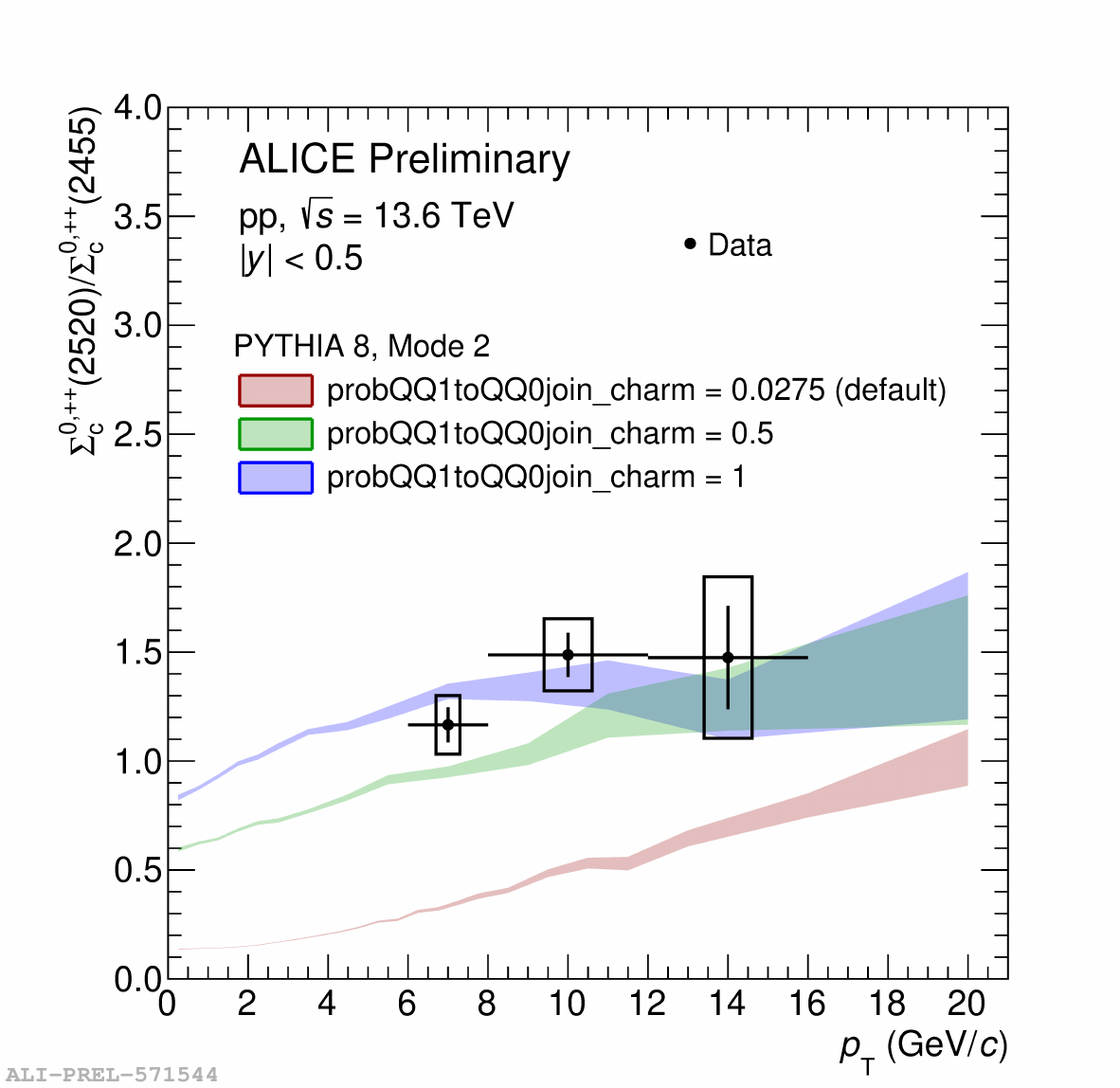}
	\caption{Relative production of the $\Sigma_{\rm c}^{0,++}(2520)$ baryon to the ground state $\Sigma_{\rm c}^{0,++}(2455)$ measured at midrapidity in pp collisions at $\sqrt{s}$ = 13.6 TeV. In the left panel, it is compared with model calculations~\cite{pythia8monash,PythiaBLC,He,SHM} and to lower energy results from Belle Collaboration (see reference on the plot). In the right panel, the ratio is compared with PYTHIA 8 CR-BLC calculations~\cite{PythiaBLC} after tuning one parameter of the model (see text for details). }
\label{Sigmac}       
\end{figure*}
\newline Moving to the HF baryon sector, the ALICE Collaboration measured at midrapidity, for the first time at the LHC, the relative production of the $\Sigma_{\rm c}^{0,++}(2520)$ baryon to the ground state $\Sigma_{\rm c}^{0,++}(2455)$. In the left panel of Fig.~\ref{Sigmac}, the ratio is compared to several model calculations~\cite{pythia8monash,PythiaBLC,He,SHM} and to similar measurements in ${\rm e}^{+}{\rm e}^{-}$ collisions at $\sqrt{s}$ = 10.52 GeV from the Belle Collaboration (see reference on the figure). No evidence of an enhancement with respect to ${\rm e}^{+}{\rm e}^{-}$ collisions is observed considering the current uncertainties. Models struggle to reproduce the data; in particular, PYTHIA 8 with the Monash tune overestimates the data, while the CR-BLC tune, which describes the prompt $\Lambda_{\rm c}^{+}$/${\rm D}^{0}$ production yield ratio fairly well, significantly underestimates the $\Sigma_{\rm c}^{0,++}(2520)$/$\Sigma_{\rm c}^{0,++}(2455)$ yield ratio. The statistical hadronisation model (SHMc)~\cite{SHM}, which assumes that the abundances of all particles, including open and hidden charm hadrons, are determined by thermal weights at the chemical freeze-out, provides the $p_{\rm T}$-integrated $\Sigma_{\rm c}^{0,++}(2520)$/$\Sigma_{\rm c}^{0,++}(2455)$ ratio which is consistent with the $p_{\rm T}$-differential measurement from ALICE. 
The statistical hadronisation model coupled with the Relativistic Quark Model (SHM+RQM)~\cite{He}, for which an augmented baryon production is obtained from the significant feed-down from excited charm baryon states beyond those listed in the PDG, predicts lower values compared to the measurement. An interesting example of how new measurements can be used to provide constraints for model tuning is shown in the right panel of Fig.~\ref{Sigmac}. In particular, the  $\Sigma_{\rm c}^{0,++}(2520)$/$\Sigma_{\rm c}^{0,++}(2455)$ production yield ratio can be reproduced by PYTHIA 8 calculations with the CR-BLC tune after the parameter that suppresses spin-one with respect to spin-zero charm-light diquark systems is tuned on the measured $\Lambda_{\rm c}^{+} (\leftarrow \Sigma_{\rm c}^{0,++})/ \Lambda_{\rm c}^{+}$ production yield ratio (see details in Ref.~\cite{Altmann}).      

\section{Highlights from p--Pb and Pb--Pb collisions}

\begin{figure*}
\centering
\includegraphics[width=4.7cm,clip]{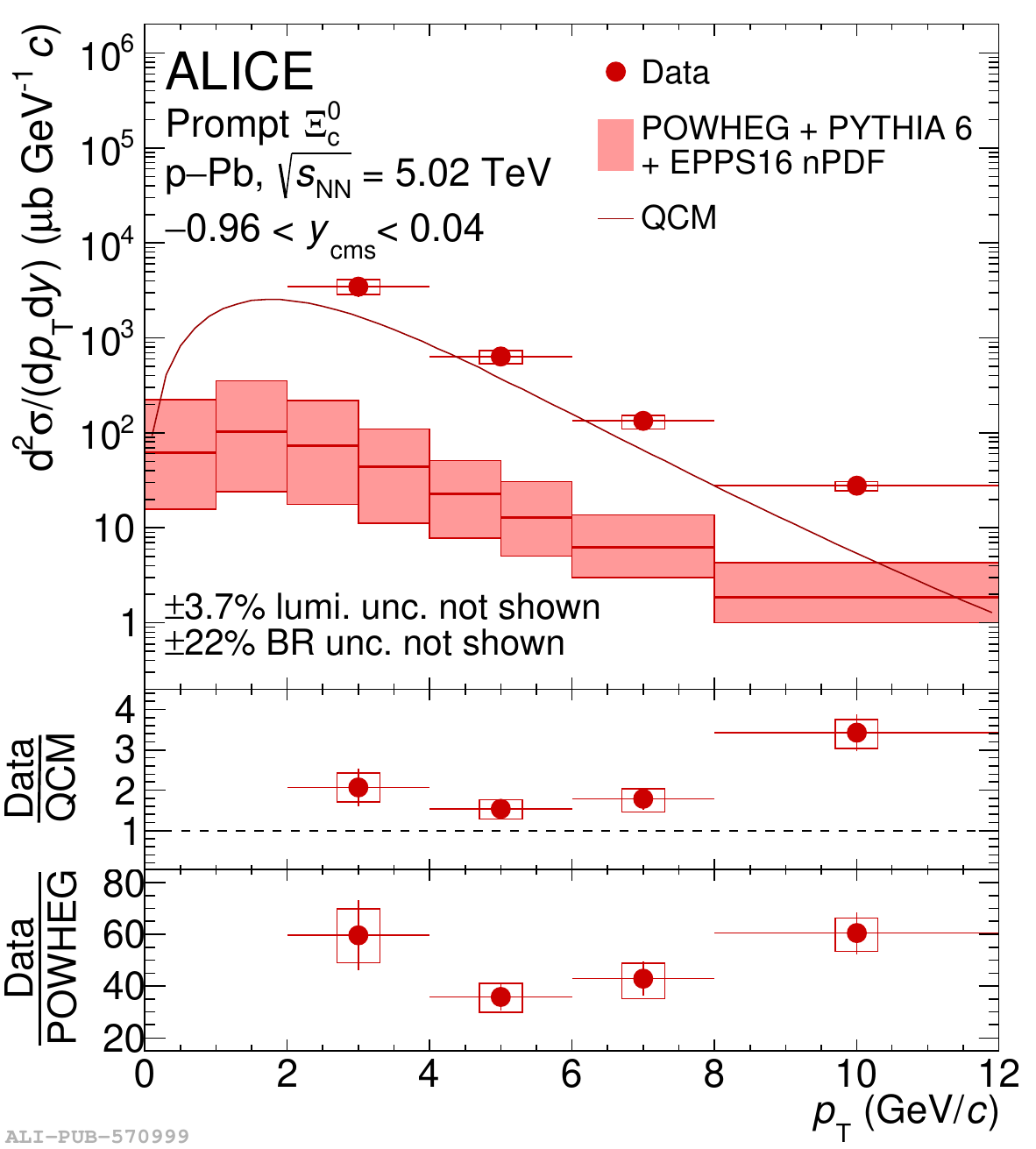}
\hspace{0.4cm}
\includegraphics[width=5.9cm,clip]{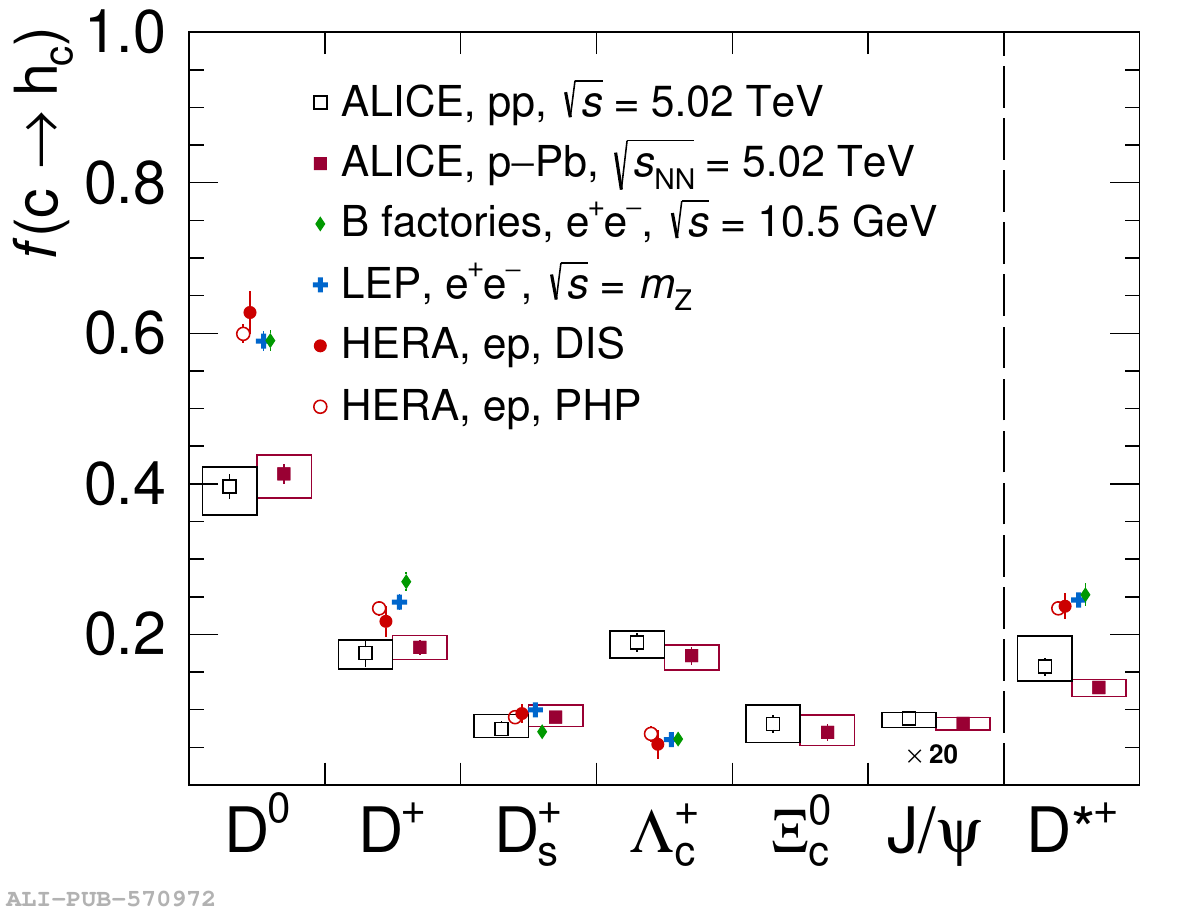}
	\caption{Left panel: prompt $\Xi_{c}^{0}$ $p_{\rm T}$-differential cross section in p--Pb collisions at $\sqrt{s_{\rm NN}}$ = 5.02 TeV compared with POWHEG+PYTHIA 6 simulations coupled to the EPPS16 nPDF parametrisation and to QCM (see~\cite{arxivPPb} and references therein). Right panel: fragmentation fractions for charm hadrons in p--Pb collisions at $\sqrt{s_{\rm NN}}$ = 5.02 TeV, compared to results from pp collisions at $\sqrt{s}$ = 5.02 TeV and from ${\rm e}^{+}{\rm e}^{-}$ and ep collisions 
	at lower energies (see~\cite{arxivPPb2} and references therein). }
\label{Xic}       
\end{figure*}

\begin{figure*}
\centering
\includegraphics[width=6.0cm,clip]{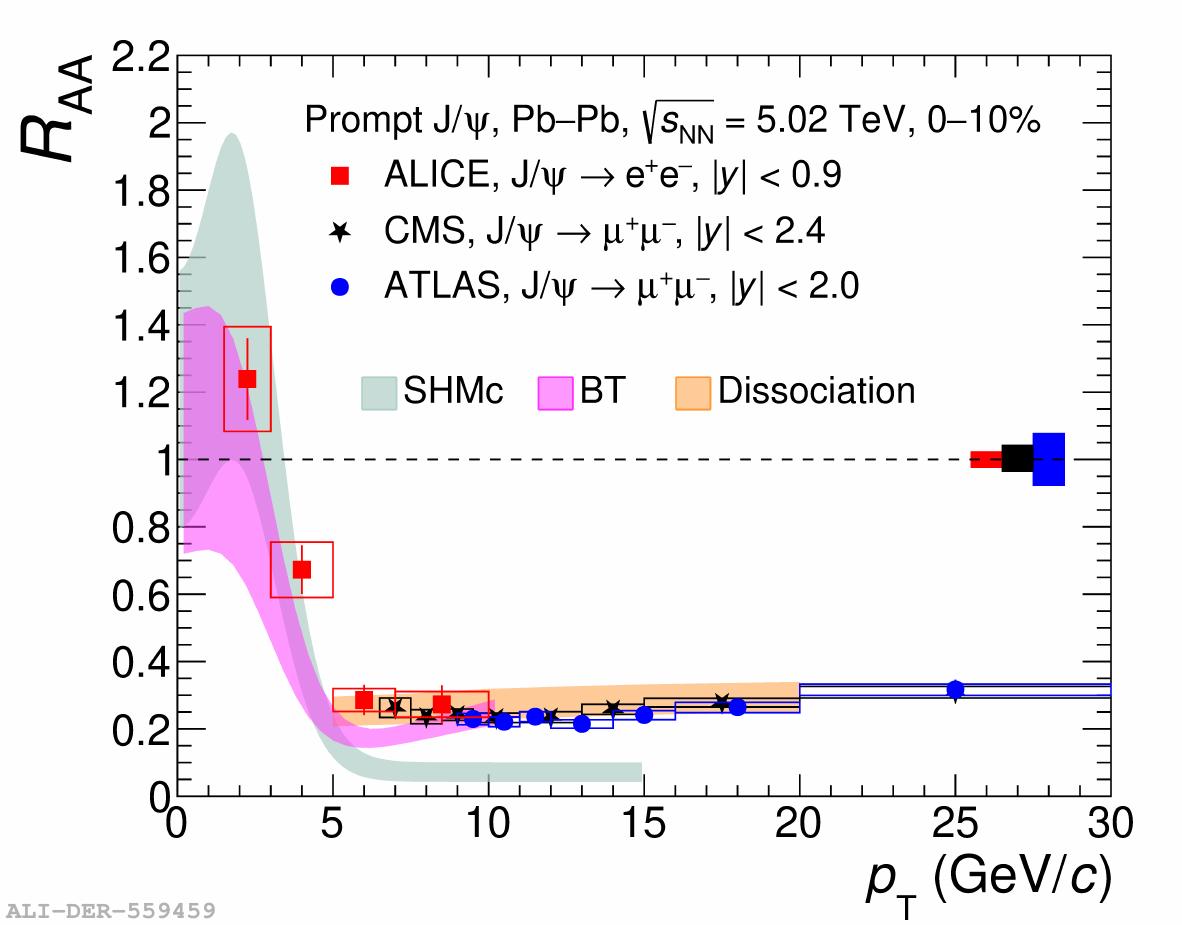}
\hspace{0.4cm}
\includegraphics[width=6.0cm,clip]{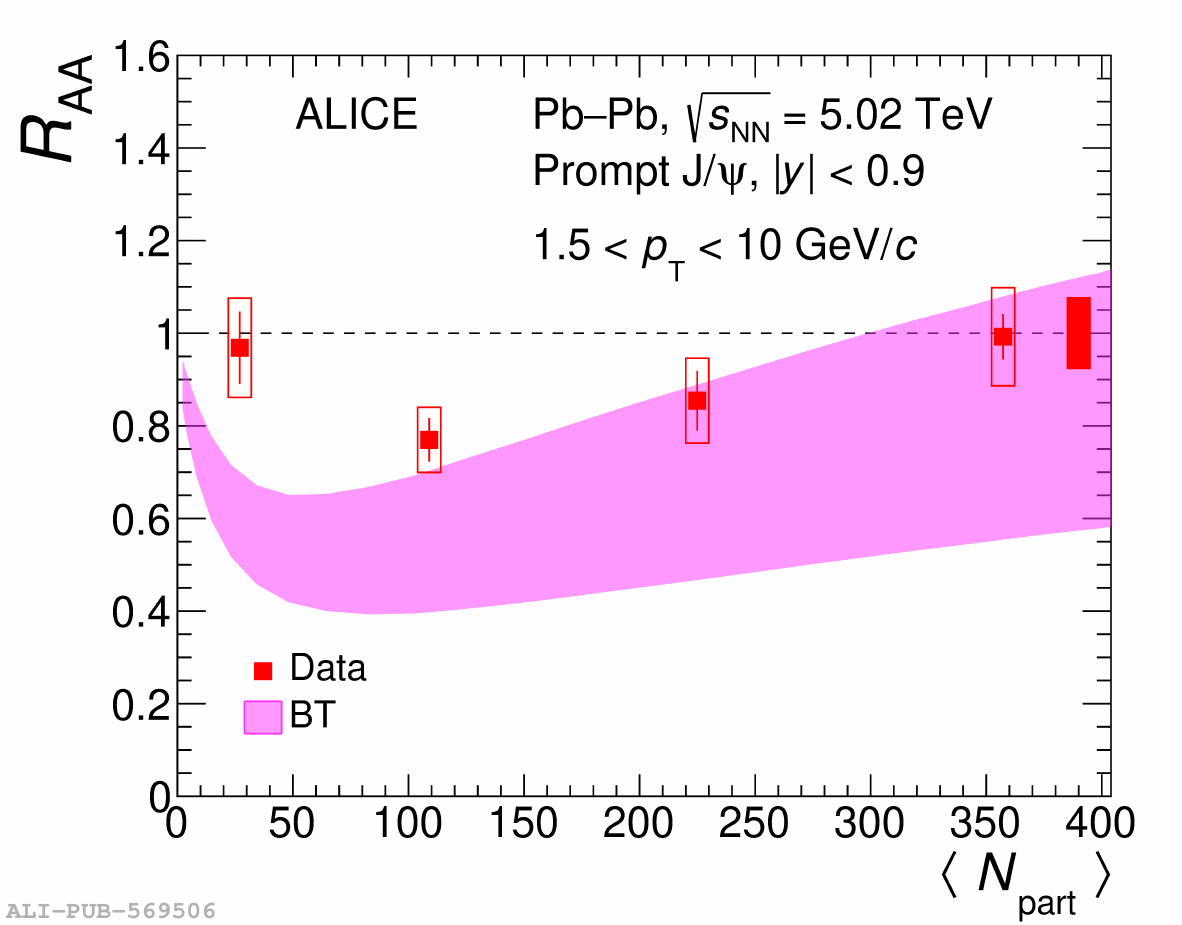}
	\caption{Left panel: $p_{\rm T}$-differential prompt J/$\psi$ $R_{\rm AA}$ measured at midrapidity in Pb--Pb collisions (0--10\% centrality class) at $\sqrt{s_{\rm NN}}$ = 5.02 TeV compared to similar measurements from other experiments and with theoretical models (see~\cite{PbPbJpsi} and references therein). Right panel: prompt J/$\psi$ $R_{\rm AA}$ for $1.5 < p_{\rm T} < 10$ GeV/$c$ as a function of $\langle N_{\rm part} \rangle$ compared with BT calculations (see~\cite{PbPbJpsi} and references therein).}
\label{JpsiRaa}       
\end{figure*}

In the left panel of Fig.~\ref{Xic}, the prompt $\Xi_{\rm c}^{0}$ $p_{\rm T}$-differential cross section in p--Pb collisions at $\sqrt{s_{\rm NN}}$ = 5.02 TeV~\cite{arxivPPb} is compared with POWHEG+PYTHIA 6 simulations coupled to the EPPS16 nPDFs (see references in~\cite{arxivPPb}). These predictions, which incorporate fragmentation fractions from ${\rm e}^{+}{\rm e}^{-}$ collisions, significantly underestimate the prompt $\Xi_{\rm c}^{0}$ production. The agreement improves with the predictions from Quark Combination Model (QCM), which assumes the coalescence of charm- and light-flavour quarks close in the momentum space, without considering the coordinate space. However, also in this case the prompt $\Xi_{\rm c}^{0}$ cross section remains underestimated by a factor of 2. The measurement of the $\Xi_{\rm c}^{0}$ production combined with the others already available for charm mesons and baryons in p--Pb collisions at $\sqrt{s_{\rm NN}}$ = 5.02 TeV, allowed for measuring fragmentation fractions of charm hadrons in this collision system. They are shown in the right panel of Fig.~\ref{Xic} where they are compared with results from pp collisions at $\sqrt{s}$ = 5.02 TeV as well as results from ${\rm e}^{+}{\rm e}^{-}$ and ep collisions at lower $\sqrt{s}$ (see~\cite{arxivPPb2} and references therein). The fragmentation fractions in p--Pb collisions are consistent with those in pp collisions, showing no dependence on the system size at the LHC. Furthermore, they are significantly enhanced (diminished) compared to those from leptonic collisions in the case of baryons (mesons), confirming the hypothesis of differing hadronisation in a parton-rich environment as already argued from previous pp measurements.   
\newline In the charmonium sector, midrapidity production measurements of prompt and non-prompt J/$\psi$ in Pb--Pb collisions at $\sqrt{s_{\rm NN}}$ = 5.02 were recently published~\cite{PbPbJpsi}. In the left panel of Fig.~\ref{JpsiRaa}, the $p_{\rm T}$-differential prompt J/$\psi$ nuclear modification factor $R_{\rm AA}$, measured in the centrality class 0--10\%, is compared to similar measurements from the CMS and ATLAS Collaborations, as well as with several model calculations (see references in ~\cite{PbPbJpsi}).  Results from ALICE nicely extend high-$p_{\rm T}$ measurements down to 1.5 GeV/$c$, and are in good agreement with CMS and ATLAS measurements in the overlapping $p_{\rm T}$ intervals. The comparison with calculations suggests that models which include re-combination either at the chemical freeze-out (SHMc) or during the QGP phase (Boltzmann Transport, BT) are able to describe the rising trend of the $R_{\rm AA}$ down to low $p_{\rm T}$. In the right panel of Fig.~\ref{JpsiRaa}, the $p_{\rm T}$-integrated ($1.5 < p_{\rm T} < 10$ GeV/$c$) prompt J/$\psi$ $R_{\rm AA}$ is shown as a function of centrality (expressed by the average number of participants, $\langle N_{\rm part} \rangle$), where it is compared to the BT model. Experimental data show a rising trend with increasing centrality, as expected from the re-generation scenario, and are well described within uncertainties by the BT model above $\langle N_{\rm part} \rangle \approx 50$.  

\section{Conclusions and outlook}
The ALICE Collaboration produced a large collection of new HF physics results based on Run 2 and fresh Run 3 data, the latter collected using the upgraded ALICE detector. 
In these proceedings, a few selected highlights from different collision systems are presented. Run 3 data taking is efficiently progressing, and will continue until the end of 2025. 
Several upgrade projects are expected to be installed during Long Shutdown 3 (2026-2028). Among others, the ITS3~\cite{its3} project foresees to replace the innermost 3 layers 
of the current ITS with truly cylindrical silicon layers, providing an improvement of the pointing resolution, crucial for the reconstruction of HF hadrons, by a factor 2 compared 
to the current ITS. These improvements, as well as the increase in terms of amount of collected data, which by the end of Run 4 is expected to be about 100 (1000) times larger for Pb--Pb (pp and p--Pb) compared to the total recorded during the Runs 1 and 2, will allow for high-precision HF measurements, including direct beauty reconstruction and the production of multi-charm baryons.

\end{document}